\documentclass{aa}  
\usepackage{txfonts}
\usepackage{graphicx}
\usepackage{natbib}
\usepackage[normalem]{ulem} 
\bibpunct{(}{)}{;}{a}{}{,} 
\usepackage[usenames]{color} 

\definecolor{MyRed}{rgb}{0.9,0.0,0.0} 
\definecolor{MyLightRed}{rgb}{1.0,0.0,0.0} 
\definecolor{MyPink}{rgb}{1.0,0.08,0.45} 
\definecolor{MyDarkBlue}{rgb}{0,0.08,0.45} 
\definecolor{MyDarkGreen}{rgb}{0,0.5,0.0} 

\newcommand{\teff}{T_{\mathrm{eff}}}
 
\newcommand{\COBOLD}{{\tt CO$^5$BOLD}}
\newcommand{\tauross}{\ensuremath{\tau_{\mathrm{Ross}}}}
\newcommand{\moh}{\ensuremath{[\mathrm{M/H}]}}

\begin{document}
\title{Three-dimensional hydrodynamical \COBOLD\ model atmospheres of red giant stars}
\subtitle{VI. First chromosphere model of a late-type giant}

\author{Sven Wedemeyer\inst{1} 
	\and 
	Ar\={u}nas Ku\v{c}inskas\inst{2}
	\and
	Jonas Klevas\inst{2}
	\and 
	Hans-G\"unter Ludwig\inst{3}
}
\offprints{sven.wedemeyer-bohm@astro.uio.no}

\institute{Institute of Theoretical Astrophysics, University of Oslo,
  P.O. Box 1029 Blindern, N-0315 Oslo, Norway
  \and 
  Institute of Theoretical Physics and Astronomy, Vilnius University, Saul\.{e}tekio al. 5, Vilnius LT-10221, Lithuania  
  \and
  ZAH, Landessternwarte K{\"o}nigstuhl,
  D-69117~Heidelberg,
  Germany
}

\date{6 January 2017; accepted 30 May 2017}

\abstract{}
{Although observational data unequivocally point out to the presence of chromospheres in red giant stars, no attempts have been made so far to model them using 3D hydrodynamical model atmospheres. 
We therefore compute an exploratory 3D hydrodynamical model atmosphere for a cool red giant in order to study the dynamical and thermodynamic properties of its chromosphere, as well as the influence of the chromosphere on its observable properties. 
}
%
{Three-dimensional radiation hydrodynamics simulations are carried out with the \COBOLD\ model atmosphere code 
for a star with the atmospheric parameters ($\teff\approx4010$\,K, $\log g=1.5$, $\moh=0.0$), which are similar to those of the K-type giant star Aldebaran ($\alpha$~Tau).  
The computational domain extends from the upper convection zone into the chromosphere ($7.4\geq\log\tauross\geq -12.8$) and covers several granules in each horizontal direction.
Using this model atmosphere, we compute the emergent continuum intensity maps at different wavelengths, spectral line profiles of \ion{Ca}{ii} K, the \ion{Ca}{ii} infrared triplet line at 854.2\,nm, and H$\alpha$, as well as the spectral energy distribution (SED) of the emergent radiative flux.
}
{The initial model quickly develops a dynamical chromosphere that is characterised by propagating and interacting shock waves. 
The peak temperatures in the chromospheric shock fronts reach values on the order of up
to 5\,000\,K although the shock fronts remain quite narrow. 
Like for the Sun, the gas temperature distribution in the upper layers is composed of a cool component due to adiabatic cooling in the expanding post-shock regions and a hot component due to shock waves. 
For this red giant model, the hot component is a rather flat high-temperature tail, which nevertheless affects the resulting average temperatures significantly. 
}
{The simulations show that the atmospheres of red giant stars are dynamic and intermittent.
Consequently, many observable properties cannot be reproduced with one-dimensional static models but demand for advanced 3D hydrodynamical modelling.  
Furthermore, including a chromosphere in the models might produce significant contributions to the emergent UV flux.
}


\maketitle

\section{Introduction}\label{sec:intro}

The existence of chromospheres around red giant stars is suggested by a  variety of observational facts. 
Mostly, the evidence comes from indirect indicators related to stellar activity, such as, for example, measurements of (variable) flux in the emission cores of \ion{Ca}{II}~H~\&~K and \ion{Mg}{II}~h~\&~k lines, asymmetries and shifts of the H$\alpha$ line core \citep[e.g.,][]{MDS08,VMC11}. 
Observations of red giants in the (sub)millimetre and centimetre ranges also point out to the presence of chromospheres \citep[e.g.,][]{DBC11,HOA13}. 
From the theoretical side, however, the properties of red giant chromospheres are still relatively poorly understood. 
For example, 1D semi-empirical models have been constructed to account for various observable properties of these stars, such as the strengths of chromospheric spectral lines, fluxes in the milimeter wavelength range, and so on \citep[e.g.,][]{mcmurry99}. 
However, these models suffer from various shortcomings. 
For instance, earlier models were unable to account simultaneously for the required ultraviolet (UV) pumping of carbon monoxide (CO) resonance lines and cool excitation temperatures of these lines in the chromosphere of Aldebaran \citep[][]{mcmurry00}.
The inability of models to reproduce cool features (e.g., CO lines) and UV lines, led authors to suggest model atmospheres with 
multiple components and the presence of shock waves in $\alpha$~Tau \citep{1999MNRAS.302...48M}. 
Detailed multi-component models with a realistic description of time-dependent shock waves require the utilisation of 
3D hydrodynamical model atmosphere codes. 
Unfortunately, to the best of our knowledge, red giant models with chromospheres computed with such codes have not been published  so far. 
Three-dimensional hydrodynamical models are well-suited for studying non-stationary stellar chromospheres and thus may help to better understand the complex interplay between non-stationary physical phenomena (such as, e.g., shock wave activity) that shape their structures.

In this work, we present the results of our first exploratory simulation of the red giant chromosphere performed with the 3D hydrodynamical \COBOLD\ model atmosphere package. 
The atmospheric parameters of the model are similar to those of the K-type giant star Aldebaran, $\alpha$~Tau (see Sect.~\ref{sec:model} below). 
Our main goal is to study the dynamical and thermodynamic structure of the chromosphere, and to investigate its influence on the observable properties of this red giant.
It must be emphasised that the simulations presented here are only a first step. 
The included physical effects restrict to what extent the model results can be interpreted. 
The exploratory model reproduces conditions as they should only be expected in rough approximation for very quiet regions in the low chromosphere. 
While this is already of interest,  $\alpha$~Tau exhibits both \ion{C}{IV} and \ion{O}{VI} features  
\citep{2005ApJ...622..629D,1998ApJ...503..396R}. 
These observations imply significantly higher temperatures than those achieved with shock heating in our 3D hydrodynamical model atmosphere and hints at yet-to-be-included physical (heating) mechanisms.

The paper is structured as follows: after a brief description of the numerical codes and models in Sect.~\ref{sec:sim}, we present the obtained results in Sect.~\ref{sec:result}, followed by a discussion and conclusions in Sects.~\ref{sec:discus} and \ref{sec:conc}. 

\begin{figure*}[t]
\includegraphics[width=\textwidth]{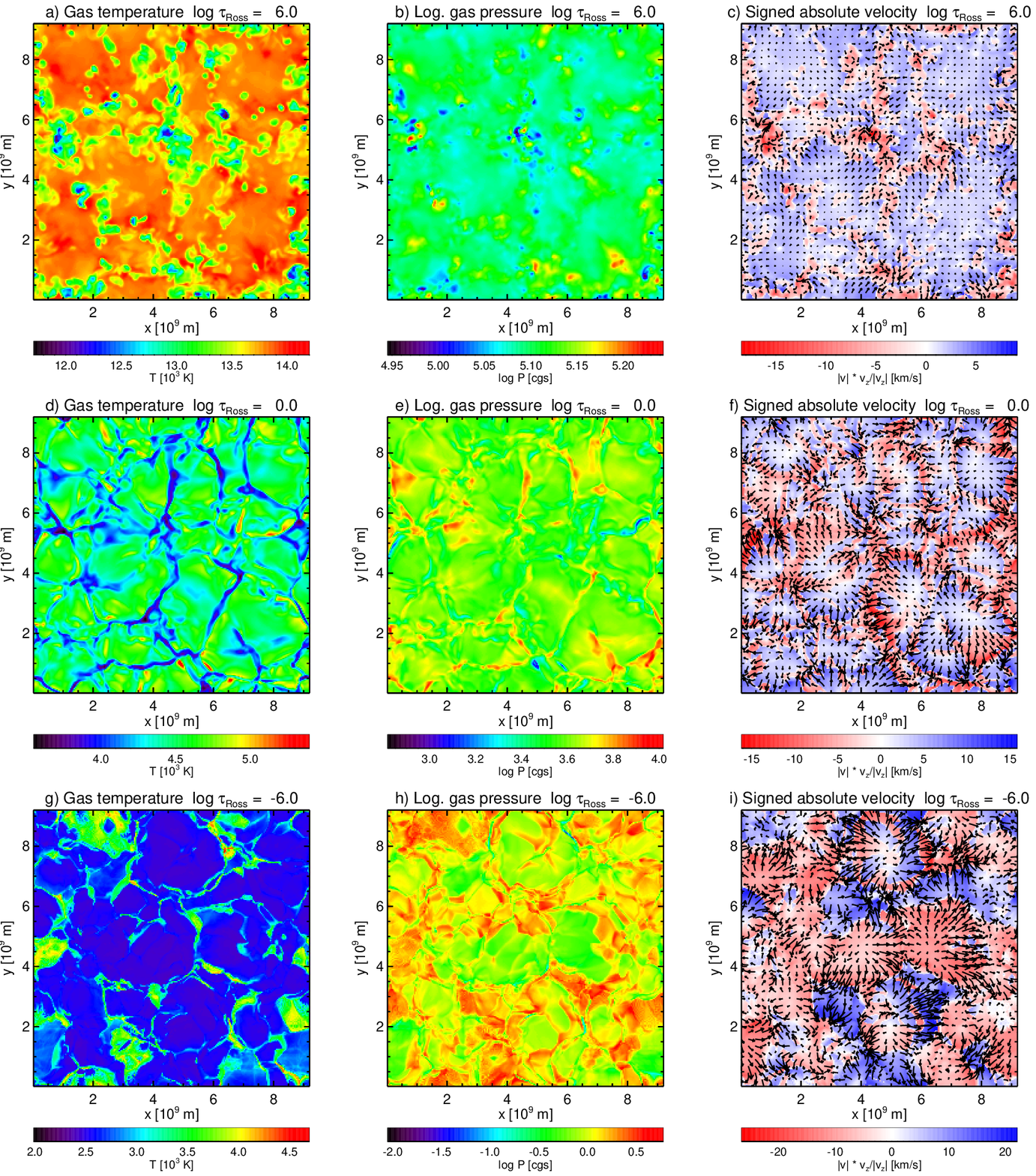}
\caption{Selected physical quantities in horizontal cross-sections through a single model snapshot (i.e., 3D model structure computed at a single instant in time) of the red giant model atmosphere after 146.8~hours simulated time. 
The different columns show maps for the gas temperature (left), logarithm of gas pressure (middle), and the product of the absolute  velocity and the sign of the vertical velocity component (right). 
Negative velocities are directed downwards/inwards (i.e., towards the stellar core) and positive velocities upwards/outwards. 
Each of these quantities is presented on planes of constant optical depth 
in the convection zone (top, $\log \tauross = 6.0$), the photosphere (middle, $\log \tauross = 0.0$), and the chromosphere (bottom, $\log \tauross = -6.0$). 
The streamlines in the right column follow the horizontal velocity field in the depicted plane. 
}
\label{fig:xyslices}
\end{figure*}

\begin{figure*}[tp!]
\centering
\includegraphics[width=\textwidth]{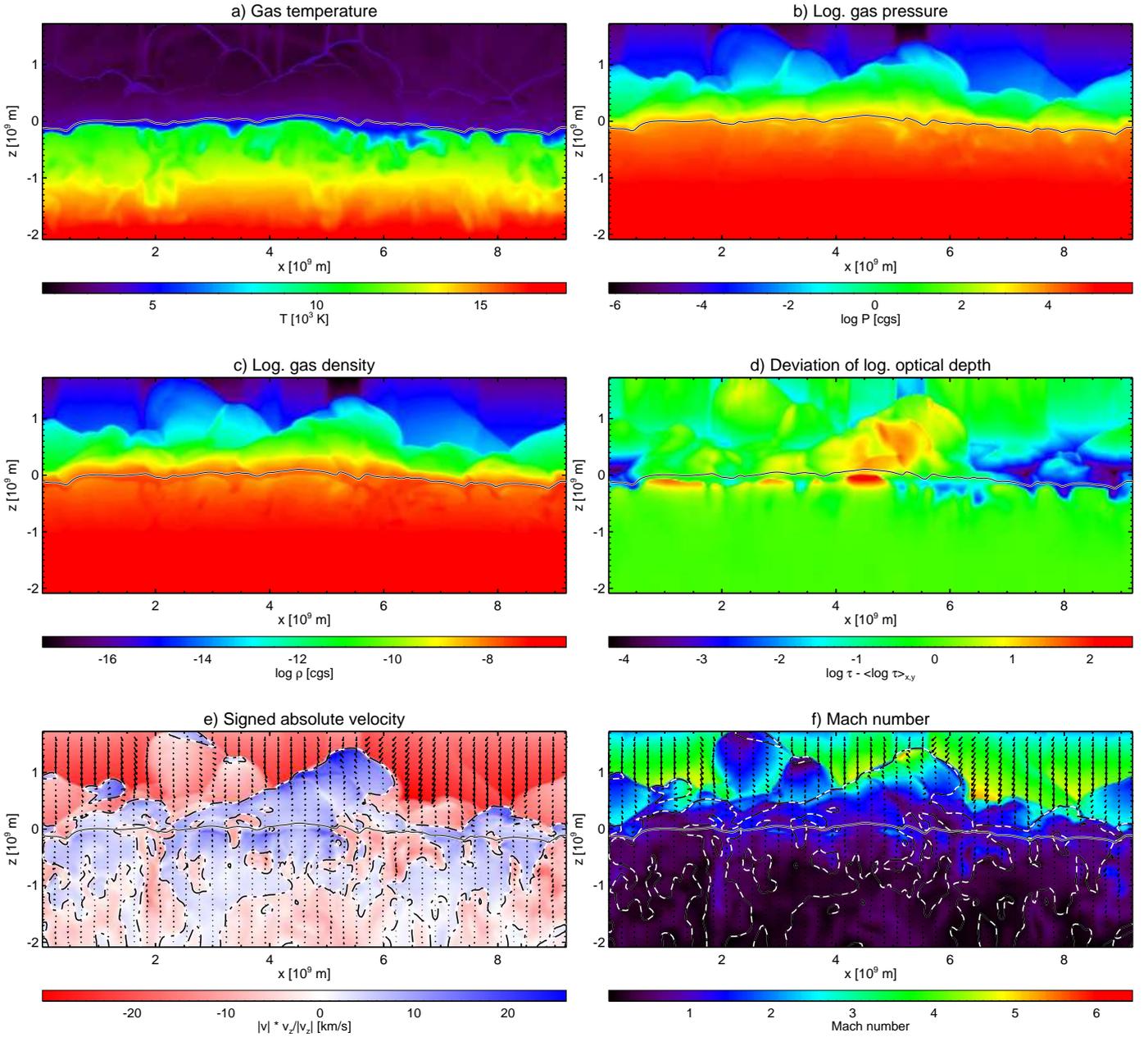}
\caption{Vertical cross-sections along the $x$-axis at $y = 0.577\cdot10^9$\,m through the same model snapshot as in  Fig.~\ref{fig:xyslices}, displaying the following physical quantities:
\textbf{a)}~gas temperature, 
\textbf{b)}~logarithmic gas pressure,  
\textbf{c)}~logarithmic gas density,  
\textbf{d)}~the difference between the local logarithmic optical depth and  its horizontal average ($\log \tauross (x,z) - <\log \tauross>_{x,y}$)
\textbf{e))}~absolute velocity multiplied by the sign of the vertical velocity component, and 
\textbf{f)}~the Mach number. 
In panels~e and d, streamlines for the flow field in the view plane and a dot-dashed contour for $v = 0$\,km\,s$^{-1}$, which thus divide upflows ($v>0$) from downflows ($v<0$), are plotted.  
The height of optical depth unity is represented by the black solid line 
around $z = 0$\,m, which thus marks the boundary between photosphere and 
convection zone. 
}
\label{fig:xzslices}
\end{figure*}

\begin{figure}[t!]
\centering
\includegraphics[width=\columnwidth]{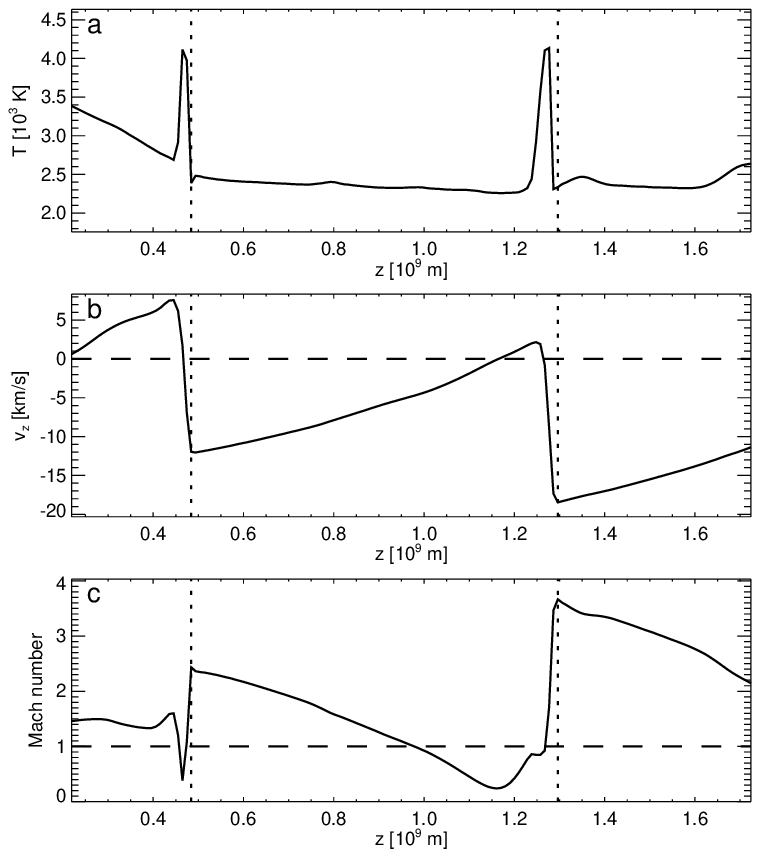}
\caption{Vertical cross-section along the height axis $z$ through the same model snapshot as in  Figs.~\ref{fig:xyslices} and \ref{fig:xzslices} at the horizontal position $[x,y] = [3.31\cdot10^9\,\mathrm{m}, 0.58\cdot10^9\,\mathrm{m}]$.  
The following physical quantities are displayed: 
\textbf{a)}~gas temperature, 
\textbf{b)}~vertical velocity, and 
\textbf{c)}~Mach number. 
The dotted vertical lines mark the positions in two major shock fronts, where the shocks meet downfalling gas.  
}
\label{fig:zprof}
\end{figure}

\section{Methodology}
\label{sec:sim}

\subsection{Three-dimensional radiation hydrodynamic simulations} 
\label{sec:cobold}

The numerical simulations are carried out with the radiation hydrodynamics  
code \COBOLD\ \citep{freytag12}. 
The hydrodynamic equations and the frequency-dependent radiative transfer equations are solved time-dependently for a fully compressible plasma in a constant gravitational field. 
The lateral boundaries are periodic. 
The bottom boundary is open so that material can flow into and out of the 
computational box. 
The entropy of the inflowing material is prescribed and sets  
the effective temperature of the simulated star. 
The top boundary is transmitting, i.e., material can leave and enter the 
model at the top. 
The equation of state is used in form of a look-up table, which is computed 
in advance under the assumption of thermodynamic equilibrium.
For this table, partial ionisation of \element{H} and \element{He}, the formation and dissociation of ${\rm H}_2$, and a representative metal are taken into account.  
The chemical element abundances are adopted from the CIFIST project 
\citep{ludwig09,CLS11}.
Non-equilibrium effects like the time-dependence of the hydrogen ionization degree in chromospheres are not considered yet for this first exploratory model due to the high computational costs but should be included in more advanced future detailed models. 
Therefore, it has yet to be seen how much the atmospheric state for the modelled stellar type would be affected when these effects are taken into account. 
%
%
A long characteristics scheme with multi-group opacities is used for the 
solution of the radiative transfer equation \citep[see, e.g.,][for more details on the opacity binning]{LJS94}. 
The opacity data is provided in form of a look-up table with five bins, which are constructed in advance using opacities from the {\tt MARCS} model atmosphere package \citep{GEK08}.

\subsection{Numerical model} 
\label{sec:model} 

The numerical model analysed here was produced with the  radiative  hydrodynamics code  \COBOLD\ (Sect.~\ref{sec:cobold} above) in several steps. 
The initial model atmosphere is based on a snapshot of a well relaxed simulation that was computed using the following atmospheric parameters: \mbox{$\teff = 4\,000$\,K}, $\log g = 1.5$, and $\moh=0.0$. This snapshot was taken from the CIFIST 3D hydrodynamical model atmosphere grid \citep[][]{ludwig09}. 
The initial model includes the top of the convection zone and the photosphere with a total extent of 
$4.6\cdot10^9\,\mathrm{m} \times 4.6\cdot10^9\,\mathrm{m} \times 2.8\cdot10^9\,\mathrm{m}\  (x,y,z)$. 
The size of the initial model  was increased to cover more granules by doubling the computational box in each horizontal direction, thus producing $2\,\times\,2$  initially identical quadrants. 
A multi-scale random velocity field was added to the new model in order to accelerate the diversion of the four quadrants and thus remove the initial symmetry as fast as possible. 
The sequence was run for one million seconds of stellar simulation time until the initial symmetry was not longer visible in the granulation. 
Next, an initially homogeneous chromosphere with 130 grid layers and a vertical extent of $1.31\cdot10^9$\,m was added on top. 
The chromosphere evolves much faster than the layers below and is therefore only added in the last step in order to save computational time. 
The final model consists of $280 \times 280 \times 280$ grid cells and has a total spatial extent of $9.2\cdot10^9\,\mathrm{m}\,\times\,9.2\cdot 10^9\,\mathrm{m}\,\times\,4.1\cdot 10^9\,\mathrm{m}$. 
The grid spacing is constant in horizontal direction ($\Delta x = \Delta y = 3.29\cdot10^7$\,m) while the vertical grid cell size decreases from $\Delta z = 5.73\cdot10^7$\,m) at the lower boundary in the convection zone to 
$\Delta z = 9.66\cdot10^6$\,m in the atmosphere, i.e. for all heights above $z = 0$\,m (except for a slight increase in the topmost layers). 
The simulation with a chromosphere is then advanced for $5.7\cdot10^5$\,sec (157~hours) of stellar time. 
The model maintains an effective temperature of $T_\mathrm{eff} \approx (4010 \pm 13)$\,K.  
For comparison, we also use a model without a chromosphere, i.e., the relaxed model obtained before adding the extra 130 grid layers. 
Note that the entropy flux at the bottom boundary of the chromospheric model was adjusted to obtain the same total emergent radiative flux as that of the model without the chromosphere. 
In effect, this ensured that the effective temperatures of the two models are nearly identical.

\subsection{Intensity synthesis} 
\label{sec:methodintensitysynthesis}

The radiative transfer code {\tt LINFOR3D}\footnote{{\tt http://www.aip.de/Members/msteffen/linfor3d}} is used to calculate the emergent continuum intensity at different wavelengths ranging from the ultraviolet to the millimeter range (see Sect.~\ref{sec:contintensity} for the results).  
{\tt LINFOR3D} is originally based on the Kiel code {\tt LINFOR}/{\tt LINLTE} and solves the radiative transfer equation in detail in 3D for an input atmosphere under the assumption of local thermodynamic equilibrium (LTE).

The same model was used as input for the {\tt MULTI\_3D} code, i.e. the column-by-column version of the original {\tt MULTI\_3D} code  \citep{carlsson1986}, which provides the detailed solution of the radiative transfer equation in non-LTE (non-local thermodynamic equilibrium).
The output contains intensity cubes ($I = f (x, y, \lambda)$) for several spectral lines of hydrogen
(e.g., H$\alpha$ and Lyman~$\alpha$)
and singly ionized calcium  (e.g., Ca\,II\,H, K, and the infrared triplet) for each selected snapshot.

\subsection{Computation of the spectral energy distributions} 
\label{sec:flux} 

The spectral energy distribution (SED) of the emergent radiation field is calculated using the {\tt NLTE3D} code \citep[][]{SPC15}. 
The code is designed to compute departure coefficients for the different energy levels of a given model atom\footnote{The departure coefficient, $b_{\rm i}$, of atomic level $i$ is defined as the ratio of population number densities, $b_{\rm i}=n_{\rm i}^{\rm NLTE}/n_{\rm i}^{\rm LTE}$, obtained under the assumptions of non-local thermodynamic equilibrium, NLTE, and local thermodynamic equilibrium, LTE.}. 
For this purpose, {\tt NLTE3D} calculates mean intensities at each grid point,  which we then use to compute the SED for the given model atmosphere, i.e., the radiative flux at its top, at different wavelengths. 

\begin{figure*}[t]
	\begin{center}
		\includegraphics[width=11cm]{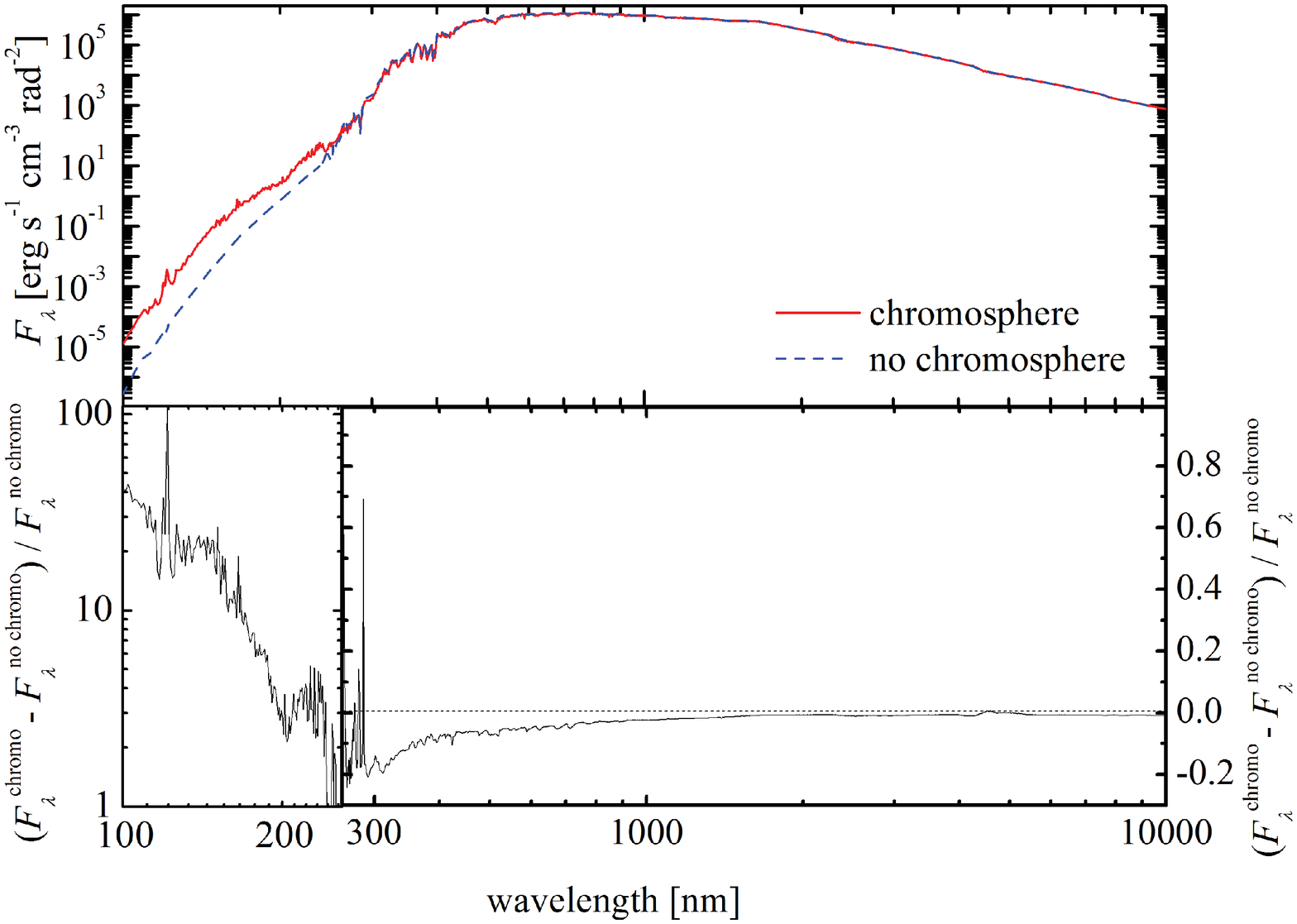}
		\caption{\textbf{Top:} Spectral energy distributions (SEDs) of the red giant models with and without chromosphere. 
		The blue line shows the average SED of the model without chromosphere computed using an ensemble of 20~model snapshots obtained at different instants in time.  
		The red line shows the SED of the model with chromosphere computed using the single snaphot shown in Fig.~\ref{fig:xyslices}.
		\textbf{Bottom:} Relative difference between the flux of the models with and without chromosphere. 
		Note that the vertical scale is different in the wavelength region $<270$~nm).}
		\label{fig:seds}
	\end{center}
\end{figure*}

We compute SEDs for two model atmospheres, one with the chromosphere and one without it (see Sect.~\ref{sec:model}). 
Note that in both cases the two models shared identical atmospheric parameters, chemical composition, opacities, and equation of state. 
The SEDs were computed in the $150-10000$~nm wavelength range. 
To account for the line opacities, we use {\tt LITTLE} opacity distribution functions (ODFs) from the {\tt ATLAS9} model atmosphere package \citep{CK03}.
Continuum opacities are taken into account by using the \COBOLD\ {\tt IONOPA} package \citep[see][for details]{SPC15}. 
The SEDs computed using the two model atmospheres are shown in Fig.~\ref{fig:seds}.

\begin{figure*}[t]
\includegraphics[width=\textwidth]{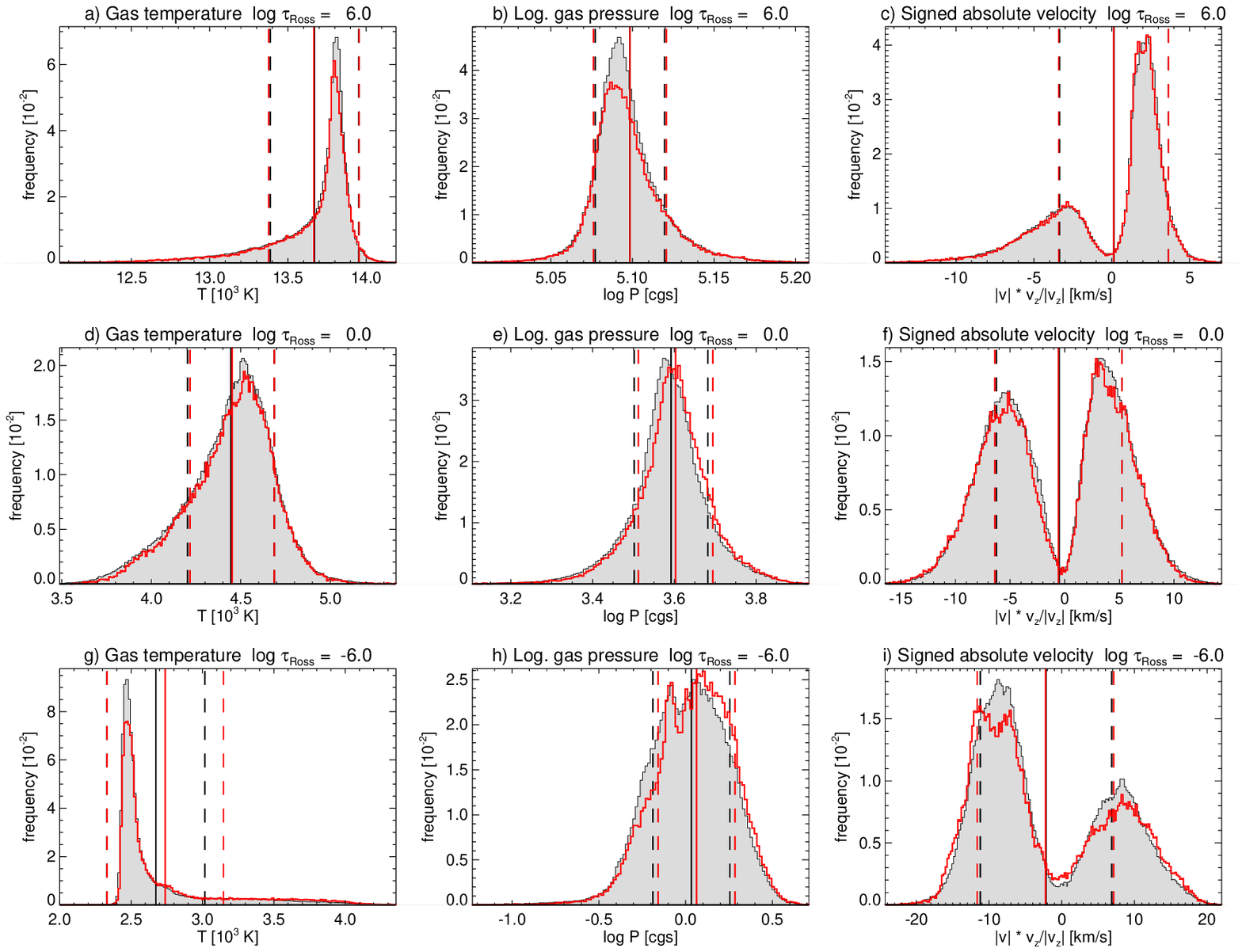}
\caption{Histograms for the horizontal cross-sections displayed in Fig.~\ref{fig:xyslices} for the same model snapshot (red) and for snapshots stretching over the second half of the simulation (grey/black). 
The columns are in the same order as in Fig.~\ref{fig:xyslices}: 
gas temperature (left), logarithm gas pressure (middle), and the product of the absolute  velocity and the sign of the vertical velocity component (right), all on planes of equal optical depth. 
The rows are, from top to bottom: convection zone (top, $\log \tauross = 6.0$), photosphere (middle, $\log \tauross = 0.0$), and  chromosphere (bottom, $\log \tauross = -6.0$). 
The solid vertical lines mark the average values whereas the dashed lines mark the average plus/minus one standard deviation. 
}
\label{fig:histograms}
\end{figure*}

\section{Results}
\label{sec:result}

The initially homogeneous atmosphere evolves quickly towards a new dynamic equilibrium state. 
The first shock waves have propagated through the whole chromosphere and reach the upper boundary of the model after 210\,000\,s. 
From then on, the whole chromosphere  is characterised by a shock-induced dynamic pattern, which is visible in many different quantities such as, e.g., gas temperature and velocity (Fig.~\ref{fig:xzslices}).  
The resulting chromosphere exhibits many features known from earlier (non-magnetic) solar models \citep[e.g.,][]{2004A&A...414.1121W,2000ApJ...541..468S} although the details are quite different. 
Propagating shock waves act as a structuring agent and produce a hot mesh of filaments, which can be 
seen in horizontal and vertical cross-sections through the selected model snapshot presented in Fig.~\ref{fig:xyslices} and Fig.~\ref{fig:xzslices}, respectively.  
The resulting complex and dynamic topology exhibits variations on a large range of spatial scales, which 
are in general larger than in solar models but also involve scales as small as the extent of the intergranular lanes in the photosphere below. 
The thermal structure of the atmosphere and the emergent continuum intensity 
are addressed in Sect.~\ref{sec:thermalstructure} and Sect.~\ref{sec:contintensity}, 
respectively.

\subsection{Atmospheric structure and dynamics} 
\label{sec:thermalstructure}

The horizontal cross-sections for a selected model snapshot after 146.8~hours simulated time in Fig.~\ref{fig:xyslices} demonstrate how the atmospheric structure changes with height in terms of gas temperature, logarithmic gas pressure, and velocity. 
The convection zone (see the panels a-c) is characterised by mostly hot upwelling gas with a mesh of cool sinking gas. 
The cool gas is mostly concentrated in plumes that have their origin in intergranular vertices in the photosphere above.  
These plumes can also be seen as areas of lower gas pressure and in the velocity, which sometimes reaches downward speeds in excess of --15\,km\,s$^{-1}$. 
These flows can thus become supersonic and reach a Mach number of up to $\sim 2$. 
The average gas temperature at the bottom of the models is \mbox{$\sim 17,500$\,K} and decreases monotonically to the effective temperature at the height where $\log \tauross = 0$, i.e., at the transition to the photosphere.

The low photosphere (see the panels d-f in Fig.~\ref{fig:xyslices}) is characterised by a granulation pattern with hot granules, where gas is rising to the surface, and narrow intergranular lanes, where the cooled gas sinks down again into the convection zone. 
The typical size of a granule is around $1.3\,-\,1.8\cdot10^9$\,m  so that about 5\,-\,7 of such granules fit next to each other in the computational box. 
However, the distribution of granule sizes in this layer seems to be otherwise continuous with larger and smaller granules occurring, too. 
The intergranular lanes typically have widths of $0.2\,-\,0.3\cdot10^9$\,m with larger downdraft areas located at granule vertices. 
There, the conservation of momentum in the downflowing plasma results in vortex flows, which seem to be an integral part of stellar surface convection (see \citealt{2012Natur.486..505W} and \citealt{2013AN....334..137W}, for vortex flows occurring in models of the Sun and M-dwarfs, respectively).

\paragraph{Shock formation.} 
The upwards propagating wave fronts steepen into shocks above $4\cdot10^8$\,m and sometimes as low as $3\cdot10^8$\,m and reach peak temperatures on the order of up to 5\,000\,K. 
The gas in the wake of shock waves is expanding adiabatically, which results in reduced gas  temperatures as low as 2\,000\,K.
The average gas temperature in the chromosphere is about 2\,500\,K. 
The velocities in shock fronts reach typical values on the order of 10\,km\,s$^{-1}$, with more extreme values of up to more than $20$\,km\,s$^{-1}$. 
As can be seen in Fig.~\ref{fig:xzslices}f, the  corresponding Mach numbers are typically on the order of 3 to 4 but can also reach values of 6 and - in very extreme, localised cases - up to 9. 
Upward propagating wave fronts can clearly be seen in vertical cross-section in Fig.~\ref{fig:xzslices}e as regions of high upward directed velocity. 
It is also imminent that these wave fronts usually  run into material that is falling down from above at high speeds, which often exceed  $-20$\,km\,s$^{-1}$. 
The situation is also illustrated for a selected column (i.e., a single  horizontal position in the computational model box) in Fig.~\ref{fig:zprof}. 
The column is taken from the vertical $x-z$ cut shown in Fig.~\ref{fig:xzslices} at a horizontal position of $x =  3.3\cdot10^9$\,m. 
In that column, two major shock fronts can be seen at heights of approximately of $z = 0.48\cdot10^9$\,m and  $z = 1.30\cdot10^9$\,m. 
The shock fronts can clearly be seen as temperature spikes and characteristic sawtooth profiles in the vertical velocity. 
The vertical velocity and the corresponding Mach number in Fig.~\ref{fig:xzslices}c illustrate the existence of supersonic downflowing material. 
The fast downflows are caused by material that has been transported up by previous shock waves and then falls down again under the influence of gravity. 
These downflows themselves contribute to the continued formation of shocks throughout the whole model chromosphere.  
The same process has been found in earlier solar models \citep[e.g.,][]{1995ApJ...440L..29C,2004A&A...414.1121W}.

\paragraph{Gas temperature distribution.} 
An important consequence of the dynamic shock pattern in the chromosphere is that the thermal and kinetic state of the gas cannot be described well with average values only. 
In other words, the complicated and intermittent structure of the modelled atmosphere cannot be described correctly with a simple one-dimensional and static model atmosphere. 
The same is also true for the modelled convection zone. 
This result becomes obvious when looking at the histograms in Fig.~\ref{fig:histograms}. 
The displayed distributions for the same model snapshot as in Figs.~\ref{fig:xyslices} and \ref{fig:xzslices} are very similar to the distributions for time steps spanning the second half of the chromospheric simulation. 
Hereafter, we refer to the latter. 
The gas temperature distribution for the chromosphere (Fig.~\ref{fig:histograms}g) has a peak at $T_\mathrm{gas} \approx 2\,450$\,K whereas the average gas temperature is 2\,670\,K. 
The true peak of the distribution is still within one standard deviation (340\,K) from the average value but there is a tail stretching to the maximum value of 4\,600\,K, which is not well presented by the average value at all. 
The high-temperature tail is due to the narrow chromospheric shock fronts, whereas the pronounced peak of the distribution is due to the cooler post-shock regions with temperatures down to $\sim 2\,000$\,K. 
Due to the non-linear dependence of gas temperature and emergent intensity, the high-temperature tail significantly affects the average temperature derived from observed intensities as we will discuss in Sect.~\ref{sec:discus}.

\begin{figure*}[t!]
\centering
\includegraphics[width=\textwidth]{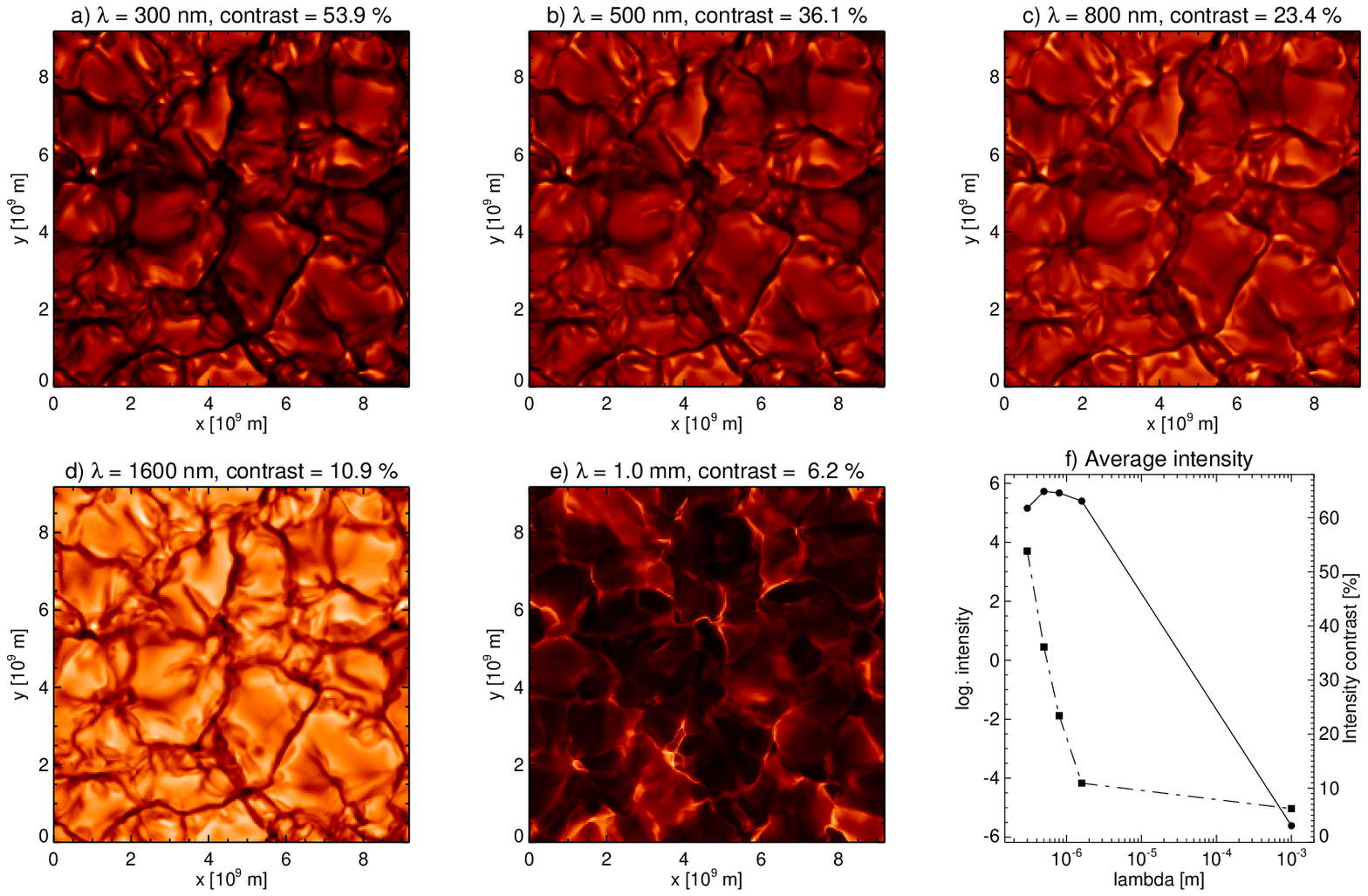}
\caption{The emergent continuum intensity for a selected simulation snapshot, at different wavelengths: 
\textbf{a)}~300\,nm, 
\textbf{b)}~500\,nm, 
\textbf{c)}~800\,nm, 
\textbf{d)}~1.6\,$\mu$m and 
\textbf{e)}~1.0\,mm. 
Panel \textbf{f)} shows the average intensity (solid line with circles) and the corresponding intensity contrast (dot-dashed line with squares) of the maps displayed in panels \textbf{a) - e)}. 
}
\label{fig:contintensity}
\end{figure*}
\begin{figure*}[tp!]
\includegraphics[width=\textwidth]{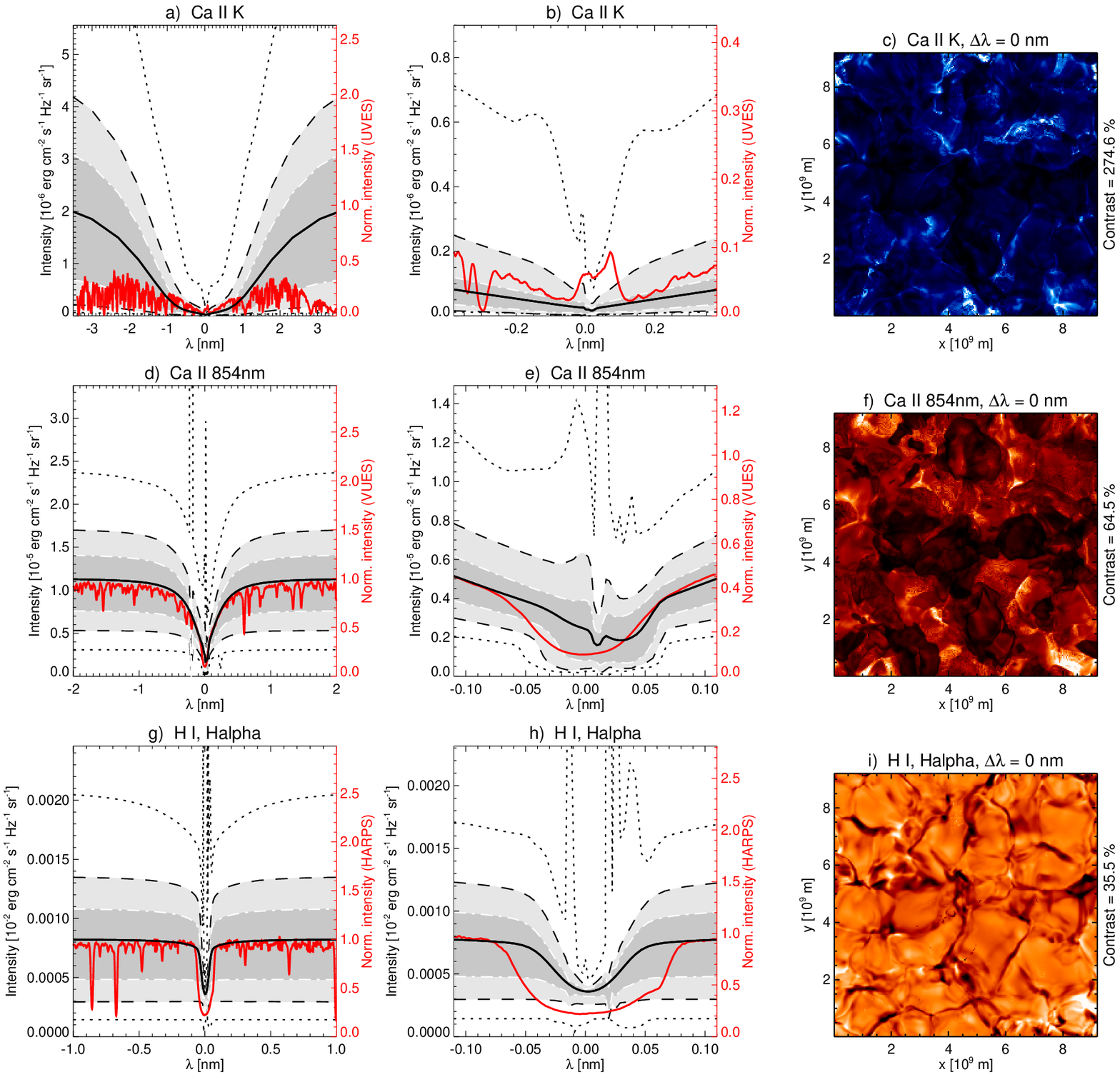}
\caption{Results of radiative transfer calculations with the \texttt{MULTI} non-LTE code for the selected model snapshot for three different spectral lines:  
Ca\,II\,K (top row), Ca\,II\,854.2\,nm (middle row), and H$\alpha$ (bottom row). 
For each row, the leftmost panel shows the horizontally averaged spectral line profile (thick solid line), the 10\,\% and 90\,\% percentiles (white dot-dashed with enclosed dark grey shaded area), the 1\,\% and 99\,\% percentiles (dashed with enclosed light grey shaded area), and the extreme values (dotted) found for that model snapshot. 
The same plots are repeated in the middle column but for a narrower wavelength range around the line core. 
For comparison, normalised observed spectra for $\alpha$~Tau are shown as red lines in the left and middle column: 
UVES POP data \citep{BJL03} in panels a-b,
VUES data in panels d-e \citep{Dob}, and 
HARPS data \citep{2014A&A...566A..98B} in panels g-h.
Note the red axis to the right of the panels. 
The horizontally resolved intensity maps for the (nominal) line core wavelength are shown in the rightmost column for a view from the top of the model, thus corresponding to stellar disk centre.}
\label{fig:multi}
\end{figure*}

\paragraph{Velocity distribution.} The velocity distributions for selected heights in the model are shown in the rightmost column of Fig.~\ref{fig:histograms}. 
Again, the distributions for the selected model snapshot (red) are very close to the distributions for the whole second half of the simulation sequence, indicating that the selected snapshot is representative. 
For all selected heights, each distribution has a pronounced peak at downward directed velocity ($v_z < 0$) and a peak for upwards directed velocity ($v_z > 0$). 
In the convection zone (see Fig.~\ref{fig:histograms}c), roughly as many grid cells with downward velocities as grid cells with upward velocities are found, resulting in an average velocity close to zero. 
However, the peak for upward velocities is narrower with a maximum around $<|v| \times   v_z/|v_z|> \approx 2$\,km\,s$^{-1}$ whereas the downward velocities span a broader range of values with a maximum close to $<|v|> \approx -3$\,km\,s$^{-1}$ and a tail reaching beyond $-10$\,km\,s$^{-1}$. 
The distribution is consistent with the rather smoothly upwelling hot gas and the cool gas shooting back into the convection zone in plumes. 
The same is in principle true for the ``surface'' layer at optical depth unity ($\log \tau_\mathrm{Ross} = 0.0$, see Fig.~\ref{fig:histograms}f) although the distribution is more symmetric in the sense that the downward velocity peak is similar (but not identical) to the upward velocity part. 
The downward velocity distribution peaks around $<|v|> \approx -5 ... -6$\,km\,s$^{-1}$ and has a tail extending well beyond $-10$\,km\,s$^{-1}$, even reaching $-15$\,km\,s$^{-1}$ in extreme cases. 
The upward velocity distribution is slightly narrower, peaks around $<|v|> \approx +3 ... +4$\,km\,s$^{-1}$ and has a tail extending well beyond $+10$\,km\,s$^{-1}$.

The situation is reversed in the chromosphere ($\log \tau_\mathrm{Ross} = -6.0$, see Fig.~\ref{fig:histograms}i) compared to the convection zone. 
In the chromosphere, slightly more grid cells exhibit downward velocities due to material falling down after having been lifted up by previous shock wave trains. 
The downward velocity part of the distribution peaks around  $<|v|> \approx -10 ... -7$\,km\,s$^{-1}$ with extreme values reaching $-20$\,km\,s$^{-1}$. 
Consequently, the arithmetic velocity average is negative, namely $<|v|> \approx -2$\,km\,s$^{-1}$. 
The upward velocity part of the distribution, which is due to upwards propagating shock waves, peaks around $<|v|> \approx +7 ... +10$\,km\,s$^{-1}$ with extreme values reaching $+20$\,km\,s$^{-1}$. 
In summary, the model chromosphere is found to be highly dynamic with the matter often travelling at  supersonic speeds and the whole atmospheric structure thus varying significantly on rather short time scales. 
At a typical speed of 10\,km\,s$^{-1}$, a shock wave (or equivalently downfalling material) propagates a distance of $10^9$\,km, which is on the order of the thickness of the modelled chromospheric layer, in on the order of $10^5$\,s.

\subsection{Properties of the radiation field} 
\label{sec:rad_field}

\paragraph{Spectral energy distributions.}
%
A comparison of SEDs computed in Sect.~\ref{sec:flux} and shown in Fig.~\ref{fig:seds} reveals that the red giant model with the chromosphere emits significantly more flux in the UV ($<270$~nm) than does the model without it. This is caused by the increasing average emissivity temperature towards the lower optical depths in the chromosphere. 
Obviously, the chromosphere makes an important contribution towards the emergent flux in the UV so that the model without a chromosphere naturally produces a significantly lower UV flux than the model with a chromosphere. 
This expected finding illustrates the need for detailed models that incorporate all atmospheric layers.

\paragraph{Continuum intensity maps.}
\label{sec:contintensity}

In Figure~\ref{fig:contintensity} continuum intensity maps at different wavelengths for the same model snapshot as in Figs.~\ref{fig:xyslices} and \ref{fig:xzslices} are presented. 
All maps represent a synthetic observation at disk center ($\mu =1.0$), i.e., as if seen spatially resolved from top. 
The maps for the wavelengths 300\,nm, 500\,nm, 800\,nm, 1.6\,$\mu$m all clearly exhibit the granulation pattern in the model although mapping slightly different height ranges around $\log \tauross = 0$. 
Consequently, the contrasts of the maps decrease from  53.85\,\% at $\lambda = 300$\,nm to only 10.93\,\% at $\lambda = 1.6\,\mu$m. 
The average intensity is shown as function of wavelengths in Fig.~\ref{fig:contintensity}f. It has a peak at 500\,nm.

The continuum intensity at millimeter wavelengths is another promising way to map the chromospheric layers. 
The large diagnostic potential of the Atacama Large Millimeter/submillimeter Array (ALMA) for this purpose has been demonstrated for the Sun \citep[see][and references therein]{2016SSRv..200....1W} and other stars, e.g., $\alpha$\,Cen \citep{2015A&A...573L...4L} and Mira \citep{2015A&A...577L...4V}.    
In Fig.~\ref{fig:contintensity}e a corresponding intensity map for a wavelength of 1.0\,mm is shown. 
In contrast to the other maps at shorter wavelengths, the continuum intensity at 1.0\,mm emerges from the chromosphere. 
The map therefore exhibits a filamentary pattern with apparent cells with diameters on the order of $1-2\cdot10^9$\,m and filament widths on the order of $0.1\cdot10^9$\,m.
The pattern in intensity corresponds to the chromospheric shock fronts and cooler post-shock regions.

\paragraph{Spectral lines.}
%
In Figure~\ref{fig:multi}, the consequences of the intermittent nature of the model chromosphere are illustrated for a few commonly used spectral lines, namely Ca\,II\,K, the Ca\,II infrared triplet line at $\lambda = 854$\,nm, and H$\alpha$. 
The figure shows spatially averaged spectral line profiles and corresponding value ranges next to intensity maps for the line cores. 
Since the line cores are expected to form highest in the model chromosphere, the resulting line core maps should exhibit the intermittent structure already seen directly in the maps for gas temperature, pressure and velocity (see Fig.~\ref{fig:xyslices}g-i).  
However, the different spectral lines are sensitive to different aspects of the model chromosphere due to non-linear dependencies in the line formation processes. 
This effect is particular obvious when comparing the Ca\,II line core maps in Fig.~\ref{fig:multi}c~and~f with the continuum intensity map at $\lambda = 1.0$\,mm (Fig.~\ref{fig:contintensity}). 
The shock-induced pattern of bright filamentary threads and dark regions is as clearly seen in the mm map as in the gas temperature map. 
The Ca\,II\,K map emphasizes more the hottest filamentary parts, whereas the picture is less clear in the   
Ca\,II\,854\,nm line core map. 
The contrast, i.e., the standard deviation of the quantity divided by its average, for the chromospheric gas temperature as shown in Fig.~\ref{fig:xyslices}g is 14.9\,\%, whereas the contrasts for the Ca\,II\,K line and the Ca\,II\,854\,nm line core intensity maps are 
274.6\,\% and 64.5\,\%, respectively.  
The mm radiation provides a more direct measure of the actual gas temperatures in the model chromosphere, which can be calculated under the assumption of local thermodynamic equilibrium (LTE), whereas the Ca\,II line cores are subject to non-LTE effects and a more complicated relationship between the state of the atmospheric gas and the line core intensity.

The spectral line profiles therefore vary significantly for the different locations. 
The intensity range covered by the individual Ca\,II\,K line profiles is illustrated in Fig.~\ref{fig:multi}. 
For some horizontal positions, the line profiles exhibit Ca\,II\,K2 peaks in the core and equivalent for the Ca\,II\,854\,nm line core as can be seen from the upper 99\,\% percentile (upper dashed lines in panels b and e, respectively). 
However, such features are not as clearly visible for many other positions so  that, despite the large spatial variations, the spatially averaged line profile does not exhibit a strong central emission reversal peak. 
While this makes it difficult to deduce the existence of shock waves from averaged spectra, as they would be observed, it obviously does not mean that shock waves do not exist. 
Fitting an observed spectrum with a one-dimensional model atmosphere would in this case lead to wrong conclusions by underestimating the extent to which shock waves occur.

The last row in Fig.~\ref{fig:multi} shows the synthetic intensities for the H$\alpha$ for the same model snapshot as discussed above. 
This spectral line is commonly used as  chromospheric diagnostics and as  activity indicator. 
Observations in H$\alpha$ on the Sun typically exhibit a complicated topology with fibrils that essentially outline the magnetic field 
\citep[see, e.g., ][]{2007ApJ...655..624D}. 
The lack of such fibrils in the H$\alpha$ line core map for the red giant model shown in Fig.~\ref{fig:multi}i is not unexpected because no magnetic fields are included in the model. 
Instead, the resulting intensity maps are dominated by photospheric contributions and thus mostly show the granulation pattern, which can be seen in the continuum intensity images in Fig.~\ref{fig:contintensity}, too. 
The information that can be retrieved from analysing the H$\alpha$ line for this particular non-magnetic model is therefore quite different from the usual diagnostic use of the H$\alpha$ line.

\section{Discussion} 
\label{sec:discus}

\begin{figure}[t]
\centering
\includegraphics[width=\columnwidth]{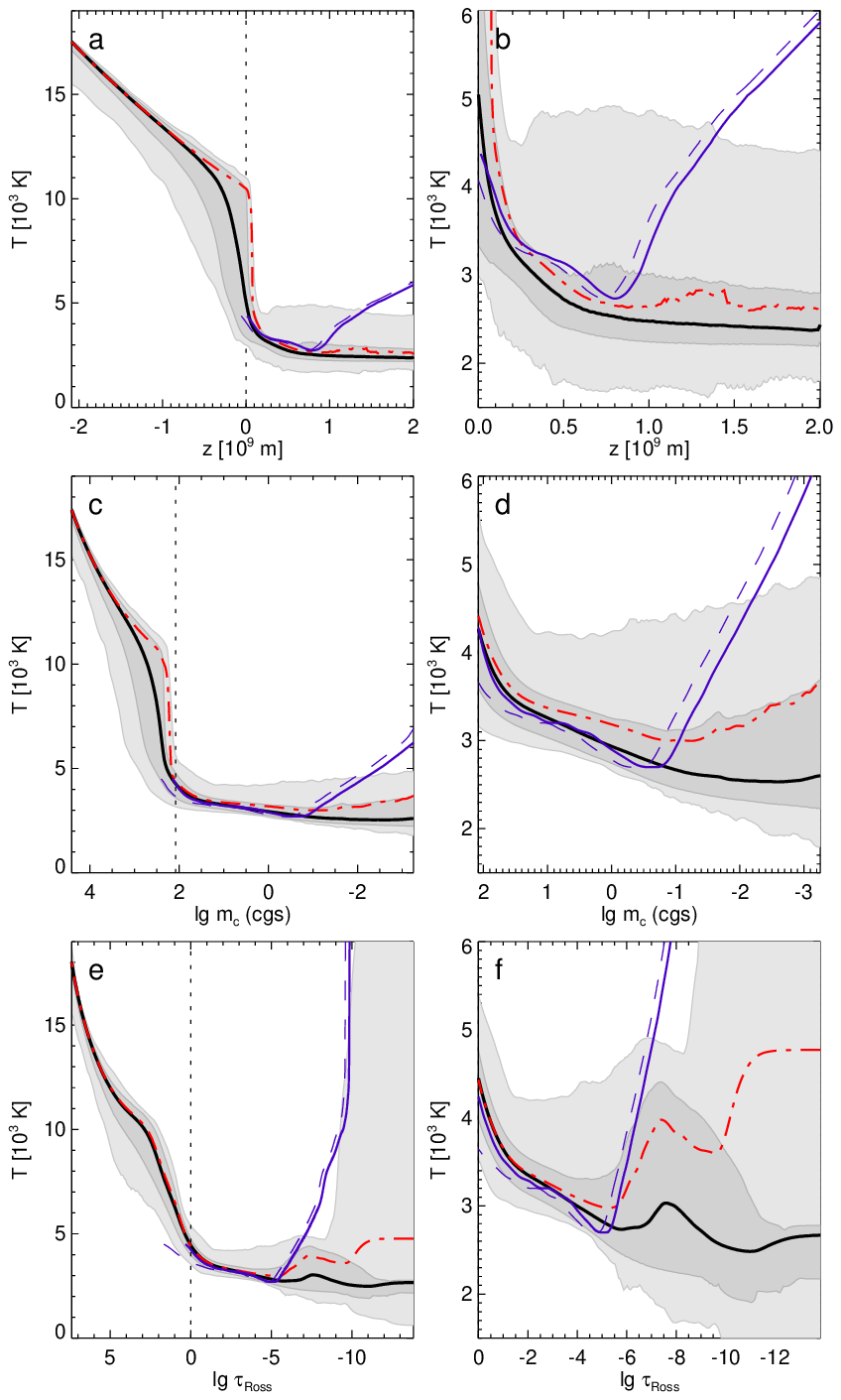}
\caption{Average stratification of the gas temperature on the \textbf{(a-b)}~geometric height scale, 
\textbf{(c-d)}~column mass scale, and 
\textbf{(e-f)}~optical depth scale. 
The right column repeats the data in the left column but for the atmospheric layers only (i.e., those located to the right of the vertical dashed line in the panels on the left). 
The arithmetic temperature averages are shown as solid black lines, whereas the red dot-dashed lines represent the corresponding average emissivity temperature as defined in (Eq.~\ref{eq:emtemp}).
The light-grey shaded areas mark the whole range of values between the minimum and maximum at each height or column mass or optical depth, whereas the dark-grey areas represent the corresponding values between the 5\,\% and 95\,\% percentiles, i.e., the majority of all values except for the 10\,\% most extreme ones. 
Please note that these are distributions for the selected model snapshot only. 
For comparison, the $\alpha$~Tau model atmosphere by \citet{mcmurry99} is plotted in all panels as blue dashed lines. 
Applying small offsets of $\Delta z = 7\,10^7$\,m and    $\Delta lg m_c = -0.25$, respectively, produces a better match with our simulation and the shifted McMurry model (thick blue line). 
The optical depth scale for the McMurry is derived from matching the column mass scale to the optical depth scale in the 3D model. 
}
\label{fig:tavg}
\end{figure}

\subsection{Preliminary comparison to observations}
%
In Figure~\ref{fig:multi}, the synthetic spectra for the presented red giant model are compared to observations of  Aldebaran ($\alpha$~Tau), namely HARPS data\footnote{The HARPS data were obtained from \texttt{http://www.blancocuaresma.com/s/}.}  \citep{2014A&A...566A..98B}, UVES and VUES \citep{Dob} data. 
A pipeline-reduced UVES spectrum (R$\sim$80\,000) was taken from the UVES POP spectral library \citep{BJL03}.
It should be noted that are remaining uncertainties regarding the scale as compared to the synthetic spectrum due to the difficulties in precisely determining the continuum in the observed spectrum. 
The VUES spectrum was obtained at Moletai Astronomical Observatory (Lithuania) using the 1.65\,m telescope and Vilnius University Echelle Spectrograph \citep[VUES][]{JFM16} in high-resolution mode (R$\sim$67\,000), taking $4 \times 60$\,sec exposures. 
The spectrum was reducing using the standard VUES pipeline procedure, and was further continuum-normalized using a synthetic 1D~LTE spectrum of Aldebaran. 
Apart from omission of the line blends in the synthetic spectra, there are substantial differences in terms of spectral line depths and widths for all three spectral lines, which are not unexpected for several reasons. 
First of all, the numerical model has an effective temperature of $T_\mathrm{eff} \approx (4010 \pm 13)$\,K, which is thus 83\,K higher than the 3927\,K for Aldebaran as stated by \citet{2014A&A...566A..98B} \citep[cf.][]{1991A&A...245..567B}. 
More importantly, the exploratory model is still lacking physical ingredients that would influence the shapes of spectral lines formed in the chromosphere. 
In particular, the numerical simulation is purely hydrodynamic and thus represents an artificially quiet state while magnetic fields have recently been detected in $\alpha$~Tau   and similar other stars 
\citep{2015A&A...574A..90A}. 
An ultimate next, although computationally expensive step, will be the inclusion of a magnetic field, non-equilibrium ionization and excitation of hydrogen and other species, which should lead to more realistic models and thus may better fit the observational data (see also Sect.~\ref{sec:magneticfield}). 
Despite this, one should also keep in mind that modeling of hydrogen line formation in red giant atmospheres is a  notoriously difficult task and that current models are consequently still not capable of reproducing observational data satisfactorily \citep[see, e.g., recent study of][and discussion therein]{2016A&A...594A.120B}. 
It may therefore still take some time until sufficiently realistic models will become available. 
Furthermore, the model presented here only includes a photosphere and a chromosphere whereas an observed spectrum naturally contains possible contributions from a corona,  circumstellar gas, incl. stellar wind \citep{2011ASPC..448.1145H}, which in principle could (partially) account for differences between the modelled and observed spectra.

It is nevertheless interesting to see that already the simplified but dynamic model chromosphere exhibits downward velocities that often exceed the upward velocities.
Consequently, the line cores are accordingly shifted for lines of sight with a strong downward velocity component, which also leaves an imprint in the averaged spectra in Fig.~\ref{fig:multi}. 
This effect is known from the solar spectrum \citep[cf.][and references therein]{1999ApJ...522.1148P}. 
A small net downward velocity on the order of $1-2$\,km/s has also been detected in the low chromosphere of $\alpha$~Tau \citep{1998ApJ...503..396R}, which may hint at the existence of shock waves.

More detailed, quantitative comparisons with observations are essential for the further improvement of numerical models and thus ultimately for a correct interpretation of the observations. 
For this purpose, complementary diagnostics that probe different aspects of stellar atmospheres should be combined. 
That includes different spectral lines and continua that are formed in different atmospheric layers and that are sensitive to different properties of the atmospheric gas (e.g., density, temperature, magnetic field).
The centre-to-limb variation as far as it can be derived from stellar observations is another important test. 
For instance, \citet{2017MNRAS.464..231R}  used lunar occultations of $\alpha$~Tau  to investigate  the limb-darkening characteristics and 
found substantial asymmetries in the photospheric brightness profile.  
Such measurements could be directly compared to corresponding synthetic observables (e.g., limb darkening profiles) as it will be presented in forthcoming publications.

\subsection{Average temperature stratification.} 
%
As mentioned above, the shock waves are clearly visible as sawtooth-like profiles in the gas temperature and velocity (Figs.~\ref{fig:xyslices}-\ref{fig:zprof}). 
The shock wave signature, however, is intermittent and thus not visible in the horizontal arithmetic average of a single snapshot over even after averaging over a few time steps.  
The average gas temperature on the geometric height scale ($z$), as shown in Fig.~\ref{fig:tavg}a-b, appears to remain roughly constant throughout the model chromosphere. 
Like for the solar models, there is no distinct temperature minimum visible in the averaged temperature stratification. 
Also, averaging on a column mass scale (Fig.~\ref{fig:tavg}c-d) or optical depth scale (Fig.~\ref{fig:tavg}e-f) produces no substantial temperature rise. 
On the other hand, the shaded areas in all panels illustrate the large range of temperature values, which is most extreme when looking at the temperature as function of optical depth in  Fig.~\ref{fig:tavg}f. 
The extended value ranges, in line with the histograms in Fig.~\ref{fig:histograms}g, strongly suggest that the chromospheric gas temperature distribution cannot be described sufficiently by means of a simple arithmetic average alone. 
In the case of an atmosphere with significant spatial and/or temporal temperature variations, deriving an average temperature stratification by adjusting a model atmosphere to reproduce an intrinsically spatially average stellar spectrum only bears limited if not potentially misleading information. 
While this statement was proven to be true for the Sun (as described below), it should be noted that the model presented here is still of exploratory nature and more advanced simulations in the future are needed for a more realistic assessment of the thermal structure of red giant chromospheres.

The non-linear dependence of the source function and thus the emergent intensity on the gas temperature, as it is for instance the case in the ultraviolet, shifts the derived average gas temperature towards high temperatures in shock fronts, which enter the average with a higher weight than the cold temperatures in the cooler post-shock regions. 
The average emissivity temperature \citep[cf.][]{2004A&A...414.1121W} 
\begin{equation} 
T_\mathrm{em} (z) = 
\left\langle {\left( \frac{\langle \kappa \rho T^4 \rangle_{x,y}}{\langle \kappa \rho \rangle_{x,y}} \right)}^{1/4}\right\rangle_{t}
\label{eq:emtemp}
\end{equation}
takes this effect into account, although rather crudely. 
The average emissivity temperature is shown as red dot-dashed lines in all panels of Fig.~\ref{fig:tavg}. 
In case of the Sun, this effect explains apparently contradicting temperature stratifications derived from UV continua and lines on the one hand and infrared carbon monoxide lines on the other hand \citep[see, e.g.,][and references therein]{1981ApJ...245.1124A,val81,1997ApJ...481..500C}.  
The high-temperature component of the chromospheric temperature distribution is less pronounced in the red giant model presented here as compared to (non-magnetic) solar models 
\citep[e.g.,][]{2004A&A...414.1121W} so that the influence of hot chromospheric gas on the average temperature is not as strong.  
Nevertheless, the emissivity temperature average produces a slightly higher average temperature in the chromosphere on the geometrical height scale (Fig.~\ref{fig:tavg}b), a notable chromospheric temperature rise on the column mass scale (Fig.~\ref{fig:tavg}d), and temperatures that are higher by up to 2\,100\,K on the optical depth scale (Fig.~\ref{fig:tavg}f). 
The resulting difference between the averages on the optical depth scale is particularly remarkable because the average emissivity temperature exceeds the 95\,\% percentile in the uppermost layers, indicating that a minority of extraordinarily hot locations in the model chromosphere strongly influence the resulting average due to the non-linear relation between the observable intensity and the actual gas temperature. 
It is this effect that has to be considered in detail when trying to derive meaningful atmospheric temperature stratification from intrinsically averaged stellar observations. 

The model atmosphere derived by \citet{mcmurry99} for $\alpha$~Tau is shown in Fig.~\ref{fig:tavg} for comparison. 
The semi-empirical, plane-parallel, hydrostatic, one-component atmosphere model was constructed to reproduce observations of optical and ultraviolet photospheric and chromospheric spectral lines including, e.g., \ion{Ca}{II}~H and K,  \ion{C}{I}, \ion{C}{II}, \ion{Si}{II}, \ion{Mg}{II} and  \ion{C}{IV}.  
There are small offsets of $\Delta z = 7\,10^7$\,m and $\Delta lg m_c = -0.25$, respectively, between the 3D numerical model and the McMurry model, which is not unexpected since the height of optical depth unity and the origin of the column mass scale vary for different horizontal positions in the 3D model so that the average scales depends on the exact way the average is calculated.  
On the column mass scale, the arithmetically averaged temperature from our simulation agrees well with the McMurry model for values of $\log m_\mathrm{c} > -1$. 
For lower values, i.e., above the low chromosphere the McMurry temperature stratification features a sharp chromospheric temperature rise and temperatures exceeding those in our model for  $\log m_\mathrm{c} < -2$ and $z > 1.4\,10^9$\,m, respectively. 
The McMurry model also includes a transition region at $z = 6\,10^9$\,m with a jump to $10^5$\,K which lies well above the upper boundary of the 3D simulation presented here. 
Although the 3D model does not exhibit a similarly strong temperature rise in the chromosphere, at least the average emissivity temperature begins to rise roughly in the same layer as the McMurry model although not as pronounced. 
As discussed above, plotting the temperatures on the optical depth scale\footnote{The optical depth scale for the McMurry model was calculated by 
converting the column mass scale with the help of the relation between optical depth and column mass in the 3D model. The resulting scale should be reasonably accurate for preliminary comparisons as presented here.} 
results in a stronger apparent chromospheric increase of the average emissivity temperature. 
The corresponding 95\,\% percentile on the optical depth scale gives the impression of a sharp temperature rise similar to the McMurry model although this behaviour only applies the hottest chromospheric grid cells in the 3D model. 
Based on the exploratory 3D model, we conclude that (i)~the chromospheric temperature rise has to be interpreted in view of potential averaging effects for a spatially intermittent atmosphere and that (ii)~the discrepancy in temperature in the upper layers implies that the 3D simulations are still lacking physical mechanisms that are important for atmospheric heating. 
Overall the situation resembles the state of development of numerical model atmospheres of the Sun in the past for which the apparent controversy regarding the chromospheric temperature structure -- in particular regarding its spatially and temporally intermittent nature -- has been largely resolved by now 
\citep[see, e.g.,][and references therein]{2008ApJS..175..229A,Gudiksen_2011AA...531A.154G,2007ASPC..368...27R,2009A&A...494..269V,2009SSRv..144..317W}.

\subsection{Magnetic fields} 
\label{sec:magneticfield}

Magnetic fields have been omitted for the first simulation presented here but will be considered for future models. 
Based on experience with \COBOLD\ models with chromospheres for the Sun 
and other stellar types \citep{freytag12,2013AN....334..137W}, we expect that the inclusion  of magnetic fields will change the chromospheric structure seen in the models significantly. 
While shock waves will also be excited in magnetohydrodynamic models, the details of their interaction with magnetic fields in the chromosphere will depend on the magnetic field strength and field topology in the model and change gradually from the non-magnetic shock-induced pattern like in the model presented here towards a magnetically dominated chromosphere with a mixture of open magnetic field lines and closed magnetic loops. 
Effects like, e.g, wave guiding, wave mode conversion, and initiation of torsional and rotational motions are expected and will not only affect the atmospheric structure but also the energy transport and thus the overall activity level of the model. 
In general, it is thought that stellar chromospheres are produced by a mixture of acoustic and magnetic heating mechanisms \citep[e.g.,][]{Narain_1996SSRv...75..453N}. 
Although the results of the red giant simulation presented in this study reveal noticeable heating of its chromospheric layers due to shock waves, the importance of magnetic heating still needs to be assessed and will be considered in our future work. 
Interestingly, although we detect vigorous shock wave activity in the chromospheric layers of the red giant modeled here, red giants that are thought  to be chromospherically inactive do not show spectroscopic signatures of shock waves \citep[e.g.,][]{1998ApJ...494..828J}. 
This finding may suggest a magnetic origin of the heating of red giant chromospheres but, on the other hand, could simply mean that the shock signatures are by nature difficult to detect due to the intrinsic averaging over the whole stellar disk as illustrated for the Ca\,II lines in Fig.~\ref{fig:multi}. 
On the other hand, there is observational evidence that suggests the presence of magnetic fields in the atmosphere of $\alpha$~Tau
\citep[e.g.,][and references therein]{2011ASPC..448.1145H,2015A&A...574A..90A}, which thus motivates the development of magnetohydrodynamic 3D simulations of this star.

\section{Conclusions}
\label{sec:conc}

The three-dimensional hydrodynamic model presented here is a first step towards more realistic model chromospheres for red giant stars that can support a meaningful in-depth analysis of stellar observations. 
Future models will include magnetic fields and eventually further physical ingredients like time-dependent hydrogen ionisation, which are important for the chromospheric gas properties and, e.g., the formation of a transition region (if existing for the modelled star) as natural upper boundary of a chromosphere.  
Despite the simplifications, the first model presented here already exhibits a very intermittent and dynamic chromosphere, which is in line with chromospheres modelled for other stellar types. 
The presented synthetic intensity maps for spectral lines and continua formed in the chromosphere clearly show that the exhibited spatial and temporal variations cannot be correctly described with a static one-dimensional model atmosphere. 
While 1D models have become very elaborate \citep[see, e.g.,][]{2016ApJ...821L...7D}, the continued development of adequate 3D (magneto-)hydrodynamical model atmospheres is ultimately needed.

\begin{acknowledgements}
We thank V.~Dobrovolskas for providing the VUES spectrum of Aldebaran prior to its publication. 
This work was supported by grant from the Research Council of Lithuania (MIP-089/2015). 
SW acknowledges support by grant from the Research Council of  Lithuania (VIZ-TYR-158).
\end{acknowledgements}

\bibliographystyle{aa}

\begin{thebibliography}{}
	
\bibitem[{{Auri{\`e}re} {et~al.}(2015){Auri{\`e}re}, {Konstantinova-Antova},
  {Charbonnel}, {Wade}, {Tsvetkova}, {Petit}, {Dintrans}, {Drake}, {Decressin},
  {Lagarde}, {Donati}, {Roudier}, {Ligni{\`e}res}, {Schr{\"o}der},
  {Landstreet}, {L{\`e}bre}, {Weiss}, \& {Zahn}}]{2015A&A...574A..90A}
{Auri{\`e}re}, M., {Konstantinova-Antova}, R., {Charbonnel}, C., {et~al.} 2015,
  \aap, 574, A90

\bibitem[{{Avrett} \& {Loeser}(2008)}]{2008ApJS..175..229A}
{Avrett}, E.~H. \& {Loeser}, R. 2008, \apjs, 175, 229

\bibitem[{{Ayres} \& {Testerman}(1981)}]{1981ApJ...245.1124A}
    {Ayres}, T.~R. \& {Testerman}, L. 1981, \apj, 245, 1124
    
\bibitem[{Bagnulo} {et al.}(2003)]{BJL03}
Bagnulo, S., Jehin, E., Ledoux, C., Cabanac, R., Melo, C., Gilmozzi, R., \& ESO Paranal Science Operations Team
2003, Messenger, 114, 10
    
\bibitem[{Bergemann} {et~al.}(2016)]{2016A&A...594A.120B} 
    Bergemann,~M., et al., 2016, A\&A, 594,
    A120
    
\bibitem[{{Blackwell} {et~al.}(1991){Blackwell}, {Lynas-Gray}, \&
  {Petford}}]{1991A&A...245..567B}
{Blackwell}, D.~E., {Lynas-Gray}, A.~E., \& {Petford}, A.~D. 1991, \aap, 245,
  567

\bibitem[{{Blanco-Cuaresma} {et~al.}(2014){Blanco-Cuaresma}, {Soubiran},
  {Jofr{\'e}}, \& {Heiter}}]{2014A&A...566A..98B}
{Blanco-Cuaresma}, S., {Soubiran}, C., {Jofr{\'e}}, P., \& {Heiter}, U. 2014,
  \aap, 566, A98

	\bibitem[{Caffau} {et al.}(2011)]{CLS11}
	Caffau,~E., Ludwig,~H.-G., Steffen,~M., et al.
	2011, SoPh, 268, 255

    \bibitem[{Carlsson(1986)}]{carlsson1986}
    Carlsson, M. 1986, A Computer Program for Solving Multi-Level Non-LTE Radiative Transfer Problems in Moving or Static Atmospheres (Uppsala Astronomical Observatory: Report No.\ 33)

    \bibitem[{{Carlsson} \& {Stein}(1995)}]{1995ApJ...440L..29C}
    {Carlsson}, M. \& {Stein}, R.~F. 1995, \apjl, 440, L29

    \bibitem[{{Carlsson} \& {Stein}(1997)}]{1997ApJ...481..500C}
    {Carlsson}, M. \& {Stein}, R.~F. 1997, \apj, 481, 500

    \bibitem[{Castelli} {et al.}(2003)]{CK03}
    Castelli, F. \&  Kurucz, R.~L.,
    2003, Proceed. of IAU Symp. 210, Modeling of Stellar Atmospheres, eds. N. Piskunov et al., poster A20 on the enclosed CD-ROM
    
    \bibitem[{Dehaes} {et al.}(2011)]{DBC11}
    Dehaes, S., Bauwens, E., Decin, L., et al.
    2011, \aap, 533, A107
    
    \bibitem[{{De Pontieu} {et~al.}(2007){De Pontieu}, {Hansteen}, {Rouppe van der Voort}, {van Noort}, \& {Carlsson}}]{2007ApJ...655..624D}
    {De Pontieu}, B., {Hansteen}, V.~H., {Rouppe van der Voort}, L., {van Noort}, M., \& {Carlsson}, M. 2007, \apj, 655, 624


\bibitem[{{Dobrovolskas} {et al.}(in prep.)}]{Dob} 
	{Dobrovolskas}, V. {et~al.} (in preparation)

    \bibitem[{{Dupree} {et~al.}(2016){Dupree}, {Avrett}, \&
    {Kurucz}}]{2016ApJ...821L...7D}
    {Dupree}, A.~K., {Avrett}, E.~H., \& {Kurucz}, R.~L. 2016, \apjl, 821, L7

\bibitem[{{Dupree} {et~al.}(2005){Dupree}, {Lobel}, {Young}, {Ake}, {Linsky},
  \& {Redfield}}]{2005ApJ...622..629D}
{Dupree}, A.~K., {Lobel}, A., {Young}, P.~R., {et~al.} 2005, \apj, 622, 629

\bibitem[Freytag et al.(2012)]{freytag12}
	Freytag, B., Steffen, M., Ludwig, H.-G., et al.
	2012, Journal of Computational Physics, 231, 919

\bibitem[{{Gudiksen} {et~al.}(2011){Gudiksen}, {Carlsson}, {Hansteen}, {Hayek},
  {Leenaarts}, \& {Mart{\'{\i}}nez-Sykora}}]{Gudiksen_2011AA...531A.154G}
{Gudiksen}, B.~V., {Carlsson}, M., {Hansteen}, V.~H., {et~al.} 2011, \aap, 531,
  A154
	
\bibitem[{Gustafsson} {et al.}(2008)]{GEK08}
	Gustafsson,~B., Edvardsson,~B., Eriksson,~K., et al.
	2008, \aap, 486, 951
	
\bibitem[{Harper} {et al.}(2013)]{HOA13}
	Harper, G.~M., O'Riain, N., \& Ayres, T.~R.
	2013, \mnras, 428, 2064

\bibitem[{{Harper} {et~al.}(2011){Harper}, {Brown}, \&
  {Redfield}}]{2011ASPC..448.1145H}
{Harper}, G.~M., {Brown}, A., \& {Redfield}, S. 2011, in Astronomical Society
  of the Pacific Conference Series, Vol. 448, 16th Cambridge Workshop on Cool
  Stars, Stellar Systems, and the Sun, ed. C.~{Johns-Krull}, M.~K. {Browning},
  \& A.~A. {West}, 1145

\bibitem[{Jurgenson} {et al.}(2016)]{JFM16}
Jurgenson, C., Fischer, D., McCracken, T., Sawyer, D., Giguere, M., Szymkowiak, A., Santoro, F., Muller, G.
2016, JAI, 550003

\bibitem[{{Judge} \& {Carpenter}(1998)}]{1998ApJ...494..828J}
    {Judge}, P.~G. \& {Carpenter}, K.~G. 1998, \apj, 494, 828

    \bibitem[{{Liseau} {et~al.}(2015){Liseau}, {Vlemmings}, {Bayo}, {Bertone},
    {Black}, {del Burgo}, {Chavez}, {Danchi}, {De la Luz}, {Eiroa}, {Ertel},
    {Fridlund}, {Justtanont}, {Krivov}, {Marshall}, {Mora}, {Montesinos},
    {Nyman}, {Olofsson}, {Sanz-Forcada}, {Th{\'e}bault}, \&
    {White}}]{2015A&A...573L...4L}
    {Liseau}, R., {Vlemmings}, W., {Bayo}, A., {et~al.} 2015, \aap, 573, L4

\bibitem[{Ludwig} {et al.}(1994)]{LJS94}
    Ludwig, H.-G., Jordan, S., \& Steffen, M.
    1994, \aap, 284, 105
    
\bibitem[Ludwig et al.(2009)]{ludwig09}
	Ludwig, H.-G., Caffau, E., Steffen, M., et al.
	2009, \memsai, 80, 711
	
\bibitem[McMurry(1999)]{mcmurry99}
	McMurry, A.~D.
	1999, \mnras, 302, 37

\bibitem[{{McMurry} {et~al.}(1999){McMurry}, {Jordan}, \&
  {Carpenter}}]{1999MNRAS.302...48M}
{McMurry}, A.~D., {Jordan}, C., \& {Carpenter}, K.~G. 1999, \mnras, 302, 48
		
\bibitem[McMurry \& Jordan(2000)]{mcmurry00}
	McMurry, A.~D., \& Jordan, C.
	2000, \mnras, 313, 423
	
\bibitem[{Meszaros} {et al.}(2008)]{MDS08}
	Meszaros, S., Dupree, A., \& Szentgyorgyi, A.
	2008, \aj, 135, 1117
	
\bibitem[{{Narain} \& {Ulmschneider}(1996)}]{Narain_1996SSRv...75..453N}
    {Narain}, U. \& {Ulmschneider}, P. 1996, Space Science Reviews, 75, 453

\bibitem[{{Peter} \& {Judge}(1999)}]{1999ApJ...522.1148P}
{Peter}, H. \& {Judge}, P.~G. 1999, \apj, 522, 1148

\bibitem[{{Rutten}(2007)}]{2007ASPC..368...27R}
{Rutten}, R.~J. 2007, in Astronomical Society of the Pacific Conference Series,
  Vol. 368, The Physics of Chromospheric Plasmas, ed. P.~{Heinzel},
  I.~{Dorotovi{\v c}}, \& R.~J. {Rutten}, 27--+

\bibitem[{{Skartlien} {et~al.}(2000){Skartlien}, {Stein}, \&
    {Nordlund}}]{2000ApJ...541..468S}
    {Skartlien}, R., {Stein}, R.~F., \& {Nordlund}, {\AA}. 2000, \apj, 541, 468

\bibitem[{{Richichi} {et~al.}(2017){Richichi}, {Dyachenko}, {Pandey}, {Sharma},
  {Tasuya}, {Balega}, {Beskakotov}, {Rastegaev}, \&
  {Dhillon}}]{2017MNRAS.464..231R}
{Richichi}, A., {Dyachenko}, V., {Pandey}, A.~K., {et~al.} 2017, \mnras, 464,
  231

\bibitem[{{Robinson} {et~al.}(1998){Robinson}, {Carpenter}, \&
  {Brown}}]{1998ApJ...503..396R}
{Robinson}, R.~D., {Carpenter}, K.~G., \& {Brown}, A. 1998, \apj, 503, 396

	\bibitem[Steffen et al.(2015)]{SPC15} 
	Steffen, M., Prakapavi{\v c}ius, D., Caffau, E., et al.\ 
	2015, \aap, 583, A57

\bibitem[{{Vecchio} {et~al.}(2009){Vecchio}, {Cauzzi}, \&
  {Reardon}}]{2009A&A...494..269V}
{Vecchio}, A., {Cauzzi}, G., \& {Reardon}, K.~P. 2009, \aap, 494, 269


    \bibitem[{{Vernazza} {et~al.}(1981){Vernazza}, {Avrett}, \& {Loeser}}]{val81}
    {Vernazza}, J.~E., {Avrett}, E.~H., \& {Loeser}, R. 1981, \apjs, 45, 635

	\bibitem[{Vieytes} {et al.}(2011)]{VMC11}
	Vieytes, M., Mauas, P., Cacciari, C., Origlia, L., \& Pancino, E.
	2011, \aap, 526, A4

    \bibitem[{{Vlemmings} {et~al.}(2015){Vlemmings}, {Ramstedt}, {O'Gorman},
    {Humphreys}, {Wittkowski}, {Baudry}, \& {Karovska}}]{2015A&A...577L...4V}
    {Vlemmings}, W.~H.~T., {Ramstedt}, S., {O'Gorman}, E., {et~al.} 2015, \aap,
    577, L4

    \bibitem[{{Wedemeyer} {et~al.}(2016){Wedemeyer}, {Bastian}, {Braj{\v s}a}, {Hudson}, {Fleishman}, {Loukitcheva}, {Fleck}, {Kontar}, {De Pontieu}, {Yagoubov}, {Tiwari}, {Soler}, {Black}, {Antolin}, {Scullion}, {Gun{\'a}r}, {Labrosse}, {Ludwig}, {Benz}, {White}, {Hauschildt}, {Doyle}, {Nakariakov},  {Ayres}, {Heinzel}, {Karlicky}, {Van Doorsselaere}, {Gary}, {Alissandrakis}, {Nindos}, {Solanki}, {Rouppe van der Voort}, {Shimojo}, {Kato}, {Zaqarashvili}, {Perez}, {Selhorst}, \& {Barta}}]{2016SSRv..200....1W}
    {Wedemeyer}, S., {Bastian}, T., {Braj{\v s}a}, R., {et~al.} 2016, \ssr, 200, 1

\bibitem[{{Wedemeyer} {et~al.}(2004){Wedemeyer}, {Freytag}, {Steffen}, {Ludwig}, \& {Holweger}}]{2004A&A...414.1121W}
    {Wedemeyer}, S., {Freytag}, B., {Steffen}, M., {Ludwig}, H.-G., \& {Holweger}, H. 2004, \aap, 414, 1121

\bibitem[{{Wedemeyer} {et~al.}(2013){Wedemeyer}, {Ludwig}, \&
      {Steiner}}]{2013AN....334..137W}
    {Wedemeyer}, S., {Ludwig}, H.-G., \& {Steiner}, O. 2013, Astronomische Nachrichten, 334, 137

\bibitem[{{Wedemeyer-B{\"o}hm} {et~al.}(2009){Wedemeyer-B{\"o}hm}, {Lagg}, \&
  {Nordlund}}]{2009SSRv..144..317W}
{Wedemeyer-B{\"o}hm}, S., {Lagg}, A., \& {Nordlund}, {\AA}. 2009, Space Science
  Reviews, 144, 317


\bibitem[{{Wedemeyer-B{\"o}hm} {et~al.}(2012){Wedemeyer-B{\"o}hm}, {Scullion},
    {Steiner}, {van der Voort}, {de La Cruz Rodriguez}, {Fedun}, \& {Erd{\'e}lyi}}]{2012Natur.486..505W}
    {Wedemeyer-B{\"o}hm}, S., {Scullion}, E., {Steiner}, O., {et~al.} 2012, \nat, 486, 505
	
\end{thebibliography}
%

%
\end{document}